\documentclass[superscriptaddress,twocolumn,showpacs,floatfix,prc]{revtex4}
\usepackage{graphicx}
\renewcommand{\vec}[1]{\mbox{\boldmath ${#1}$}}

\begin{document}

\title{
Spin-orbit induced backflow in neutron matter with
auxiliary field diffusion Monte Carlo
}

\author{L.~Brualla}
\email{lbrualla@sissa.it}
\affiliation{International School for Advanced Studies, SISSA, 
and \\ INFM {\sl DEMOCRITOS} National Simulation Center \\ Via Beirut
I-34014 Trieste, Italy}
\author{S. Fantoni}
\email{fantoni@sissa.it}
\affiliation{International School for Advanced Studies, SISSA, 
and \\ INFM {\sl DEMOCRITOS} National Simulation Center \\ Via Beirut
I-34014 Trieste, Italy}
\author{A. Sarsa}
\email{sarsa@ugr.es}
\altaffiliation{Present address: 
Departamento de F\'{\i}sica Moderna, Universidad de Granada, E-18071 Granada,
Spain}
\affiliation{International School for Advanced Studies, SISSA, 
and \\ INFM {\sl DEMOCRITOS} National Simulation Center \\ Via Beirut
I-34014 Trieste, Italy}
\author{K. E. Schmidt}
\email{kevin.schmidt@asu.edu}
\altaffiliation{Permanent address: Department of Physics and Astronomy, 
Arizona State University, Tempe, AZ, 85287}
\affiliation{International School for Advanced Studies, SISSA, 
and \\ INFM {\sl DEMOCRITOS} National Simulation Center \\ Via Beirut
I-34014 Trieste, Italy}
\author{S.~A.~Vitiello}
\email{vitiello@unicamp.br}
\affiliation{Instituto de F{\'\i}sica {\sl Gleb Wataghin}\\
Universidade Estadual de Campinas, Unicamp}

\date{\today}

\begin{abstract}
The energy per particle of zero-temperature neutron matter is investigated,
with particular emphasis on the role of the 
$\vec L\cdot\vec S$ interaction.  An analysis
of the importance of explicit spin--orbit correlations in the description
of the system is carried out by the auxiliary field diffusion Monte
Carlo method. The improved nodal structure of the
guiding function, constructed by explicitly considering these correlations,
lowers the energy. The proposed spin--backflow orbitals can conveniently
be used also in Green's Function Monte Carlo calculations of light nuclei.
\end{abstract}

\maketitle

\section{Introduction}
In recent investigations \cite{sarsa03,fantoni01b} of the ground state and 
the magnetic properties of neutron matter with
modern nuclear interactions of the 
Urbana-Argonne type \cite{wiringa95,pudliner97}
good agreement was observed between results
obtained with the Auxiliary Field Diffusion Monte Carlo
(AFDMC) \cite{schmidt99}, a calculation by 
\textcite{morales02},
performed with the  Variational Chain Summation (VCS) 
method \cite{akpa97,akmal98}, and Brueckner Hartree Fock
estimates \cite{baldo00,vidana02}. 

However, in spite of the overall agreement of the equation of state and
the spin susceptibility,
marked differences exist between the AFDMC and the VCS calculations, concerning
the contribution to the energy due to the spin--orbit component of the
two--body interaction. VCS calculations, performed with the 
Argonne $v_{18}$ \cite{wiringa95} and the Urbana-IX three--body potential 
(AU18 Hamiltonian), 
find large and negative contributions from the cluster terms $C_{LS}$
with either spin-orbit correlations in the trial function and/or the
spin--orbit potential. For instance, in correspondence 
with optimal trial functions,
$C_{LS}$ amounts to -5.8 MeV at a density $\rho=\rho_0=0.16$ fm$^{-3}$ and
-12.1 MeV at twice the same density. The AFDMC calculations of 
Ref.~\cite{sarsa03}, performed with the simplified, 
but still realistic, version of the Argonne two--body
potential, $v_{8'}$ \cite{pudliner97}, 
plus the Urbana IX three--body interaction ($AU8'$ Hamiltonian)
get energy differences $\Delta E_{LS}$
between the $AU8'$ Hamiltonians with and without  
the spin--orbit potential which are small and positive. At $\rho=\rho_0$, 
$\Delta E_{LS}= 0.2$ MeV and at $\rho=2 \rho_0$,
$\Delta E_{LS}= 1.12$ MeV. 

Such a discrepancy cannot be ascribed to differences between
the two potentials $v_{18}$ and $v_{8'}$. It is well known that they 
provide very close results for the energy per 
particle \cite{pudliner97}. In addition, this
discrepancy is confirmed by other FHNC/SOC calculations \cite{sarsa03},
performed by using the approximations of 
Ref.~\cite{wiringa88} and the $AU8'$
Hamiltonian. They give $\Delta E_{LS}=-3.7$  MeV at $\rho_0$ and
$\Delta E_{LS}=-10.1$ MeV at $2\rho_0$, with $C_{LS}$ being
$-3.9$ MeV and $-10.7$ MeV respectively.

A possible source for this
disagreement might be the use of a less than satisfactory guiding
function in the AFDMC method \footnote{
In the AFDMC method a path constraint is used to deal with the fermion
sign or phase problem. In all the calculations reported here the
real part of the guiding function, evaluated at the walker position
and spin, is constrained to be positive in analogy with the fixed
node approximation for central potentials. For convenience we refer
to the effect of the constraint as the nodal structure of the guiding
function.
}. 
The nodes of the plane wave Slater
determinant might be too poor, particularly when the interaction 
includes a spin--orbit potential, like in $AU8'$. The results of the
FHNC/SOC calculations of Ref.~\cite{sarsa03} with the $AU8'$
Hamiltonian and a trial function of the type $F_6$ (not
containing spin--orbit correlations), give values  
for $\Delta E_{LS}$ quite close to the AFDMC ones, which may confirm 
this hypothesis.

To clarify this issue we modify the guiding functions in our quantum
Monte Carlo calculations to contain explicit spin--orbit
correlations. This is efficiently done by considering
orbitals of the spin--backflow form in the
Slater determinant, as explained below.

The structure of this paper is as follows. In Sec. \ref{sec.2} we show
the guiding function used. The computational details are given in
Sec. \ref{sec.3}. The results are shown and discussed in Sec. 
\ref{sec.4}. The conclusions and perspectives of the present work
can be found in Sec. \ref{sec.5}.

\section{Spin-Orbit Induced Backflow}
\label{sec.2}

The $\vec L\cdot\vec S$ correlation in FHNC/SOC and VCS calculations takes
the form

\begin{eqnarray}
F_b(1,2) = \frac{1}{4i}f_b(r_{12})\left[ \vec r_{12}\times
\left(\vec\nabla_1-\vec\nabla_2\right) \right ] \cdot\left(\vec\sigma_1+
\vec\sigma_2\right)
\label{lscorr}
  \end{eqnarray}
where $r_{12}=|\vec r_{12}|$ and $\vec\sigma_j$ are the Pauli matrices for
the $j$th particle.

By inspection of the cluster terms of $\langle AU8'\rangle$  
at the two--body order, one
observes that the leading of the spin--orbit terms are those usually denoted
as {\sl bbc} and {\sl cbb} in FHNC/SOC 
theory \cite{pandharipande79}. In the calculations
of Ref.~\cite{sarsa03} these terms are responsible for
$\sim 80\%$ of the total
contribution from all the $C_{LS}^{(2)}$ terms in the density range
$(3/4)\rho_0 \leq \rho \leq (5/2)\rho_0$.
One can easily prove that exactly the same expressions 
of the {\sl cbb} and {\sl bbc} terms are obtained by the following simplified 
$\vec L\cdot\vec S$ correlation.

\begin{eqnarray}
\tilde{F}_b(1,2) = \frac{1}{4i}f_b(r_{12})\left[ \vec r_{12} \cdot
\left(\vec\nabla_1\times\vec\sigma_1
-\vec\nabla_2\times\vec\sigma_2\right)\right] \ .
\label{lscorrback}
  \end{eqnarray}
It is found that, at the two--body level of the FHNC/SOC theory,  
$\tilde{F}_b$ leads to an energy which is only $\sim 10\%$ different from
that obtained with $F_b$.
 
The important feature of the $\vec L\cdot\vec S$ 
correlation of Eq. (\ref{lscorrback})
for AFDMC calculations is that, similarly to the case of standard 
backflow \cite{schmidt81}, it can be implemented in quantum Monte 
Carlo simulations by substituting the plane wave orbitals of the Slater 
function with the following spin--backflow ones
\begin{eqnarray}
  \label{eq:bkf}
  & &\exp(i\vec k \cdot \vec r_j)
      \rightarrow \, \\
  & & \phi_{\vec k}(j) = \exp\left(i\vec k \cdot  \vec r_j
           +{\beta\over 2} \sum_{k\ne j} f_b(r_{jk})
           (\vec r_{jk} \times \vec k) \cdot \vec\sigma_j\right) \ , \nonumber
\end{eqnarray}
where $\beta$ is a spin--orbit strength parameter. For $\beta=1$ there 
is a direct correspondence of $\phi_{\vec k}(j)$ with
$\tilde{F}_b$. This spin--backflow ansatz can also be used
in the GFMC simulations of small nuclear 
systems \cite{pudliner97,pieper02,carlson03},
to include $\vec L\cdot\vec S$ correlations. 

We present and discuss, in the following, the results obtained in 
AFDMC simulations of neutron matter energy with the $AU8'$ Hamiltonian
and the nodal surface of the spin--backflow Slater function. We will show
that these nodes serve to lower the AFDMC energy per particle of 
neutron matter by a sizable amount, which
however is too small to solve the CVS and AFDMC discrepancy, 
particularly at higher densities.

\section{Calculation Details}
\label{sec.3}

We describe the neutron system by the non relativistic Hamiltonian

\begin{eqnarray}
H & = & T+V_2+V_3 \nonumber \\
& = &-\frac{\hbar^2}{2m}\sum_{j=1,N}\nabla^2_j + 
\sum_{j<k}v_{jk} + \sum_{j<k<l} V_{jkl} \ ,
\label{ham_ham}
\end{eqnarray}

\noindent
where $m$ is assumed to be the average of the neutron and proton masses
and $\hbar^2/m=41.47108$ MeV fm$^2$;
the two-body and three-body potentials, $v_{ij}$ and $V_{jkl}$,
are the Argonne $v'_8$ and the the Urbana IX potentials \cite{pudliner97}.
The three-body contributions to the
energy are large, particularly at high densities, and cannot be neglected
even in a survey calculation like ours.

In neutron matter this interaction can be written in terms of four components:

\begin{eqnarray}
v_{ij} &=& \sum_{p=1}^4 v_p(r_{ij}) {\cal O}^p (ij)  \\
{\cal O}^p (ij) &=& 1, \vec\sigma_i\cdot\vec\sigma_j, S_{ij}, 
\vec L\cdot\vec S \ ,
\end{eqnarray}
where $S_{ij}$ and $\vec L\cdot \vec S$ are the usual tensor and spin--orbit
operators. The functions $v_p(r_{ij})$ can be found 
on Ref.~\cite{pudliner97} and also on Ref.~\cite{sarsa03}.

The Urbana-IX three-body interaction is given by the sum

\begin{eqnarray}
V_{jkl} = V_{jkl}^{SI} + V_{jkl}^{SD},
\label{v3}
\end{eqnarray}
where $V_{jkl}^{SI}$ is a spin independent three--body short range part,
and the spin--dependent part, $V_{jkl}^{SD}$, in neutron matter, 
reduces to a sum of terms containing only two--body spin operators, 
with a form and strength that depends on the positions of three particles.
Their explicit expressions can be found in Ref.~\cite{sarsa03}.

We have also considered an interaction obtained from $AU8'$ by dropping
the spin--orbit term, and, as in Ref.~\cite{sarsa03}, it is denoted as
$AU6'$. 

\begin{figure}[tb]
\includegraphics[height=\columnwidth,angle=-90]{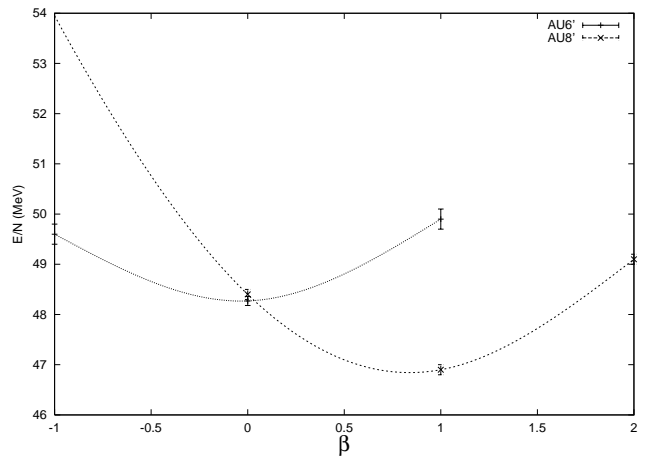}
\caption[]{Neutron matter energy per particle in MeV as a function of 
the strength $\beta$ of the spin--orbit correlation function $f_b$ at 
$\rho=2\rho_0$ for 14 neutrons.  
Solid line stands for the $AU6'$ interaction, dotted line for the 
$AU8'$ potential.
}
\label{fig1}
\end{figure}

The AFDMC \cite{schmidt99} method used in these calculations has been
previously described in detail \cite{sarsa03}.
It allows Monte Carlo simulations to be performed on a relatively large nuclear 
system at the required accuracy, thanks to the introduction of 
auxiliary fields, which 
uncouple the spin--dependent interaction between particles by means of a
Hubbard--Stratonovich transformation. While the propagation of the particle 
coordinates is done as in diffusion Monte Carlo, that of the 
spin variables, after 
having sampled the auxiliary fields, results in a rotation of each 
particle's spinor.

The guiding function used in this work is given
by a Jastrow  product correlating an $N\times N$ Slater determinant  

\begin{eqnarray}
\langle R,S |\Psi_{JSB} \rangle = \prod_{j<k} f(r_{jk}) \Phi_B(R,S) ,
\label{jastrow}
\end{eqnarray}

\noindent
where $R$ and $S$ denote the set of particle coordinates and of spinors
respectively. The space and spin orbitals in $\Phi_B(R,S)$ are given 
in Eq. (\ref{eq:bkf}), and they operate on the spin states in the 
following way 

\begin{eqnarray}
\phi_{\vec k}(i) |\uparrow\rangle &=& e^{i\vec k\cdot \vec r_i} \nonumber \\
&\times& \left[\left(\text{cosh}(A_{\vec k}(i))+\hat{A}_{\vec k}^z(i) 
\text{sinh}(A_{\vec k}(i))\right) |\uparrow\rangle\right. \nonumber \\  
&+&\left. \left(\hat{A}_{\vec k}^x(i)+i \hat{A}_{\vec k}^y(i)\right)
\text{sinh}(A_{\vec k}(i)) |\downarrow\rangle\right] \ ,
  \end{eqnarray}

\noindent
where

\begin{eqnarray}
\vec A_{\vec k}(i) = \frac{\beta}{2}\sum_{j\ne i} f_b(r_{ij}) \vec r_{ij}
\times \vec k
  \end{eqnarray}
and $\hat A_{\vec k}^{\alpha}(i)$, $\{\alpha=x,y,z\}$ are the components
of an unit vector in the direction of $\vec A_{\vec k}(i)$. 
A similar expression can be obtained for the action over the spin down 
single particle state.
We denote this function by JSB as opposed to JS which stands for a function
without explicit backflow correlations, {\sl i.~e.},
with simple plane waves in the spatial part of the orbitals in the
Slater determinant ($f_b(r)=0$). The AFDMC method described in 
Ref.~\cite{sarsa03} needs only slight modifications to deal with
$\langle R,S | \Psi_{JSB}\rangle$ as guiding function. They refer mainly to
calculating the kinetic energy part of the local energy, and the
gradient of the guiding function in the drift of the walker.

The Jastrow and spin--orbit correlation functions $f(r)$ and $f_b(r)$ 
have been taken as the first and fourth components respectively
of the FHNC/SOC correlation operator that minimizes
the energy per particle of neutron matter at the desired 
density \cite{wiringa88}. 

\section{Results}
\label{sec.4}

We have made our calculations within the full simulation box, and we 
have taken into account the 26 neighboring boxes in the tabulation of
the correlations $f(r)$ and $f_b(r)$, and of the various 
components $v_p(r)$ of the two--body potential, as described in 
Ref.~\cite{sarsa03}. Our results are therefore already tail corrected
for a Hamiltonian with two--body force only. 
Tail corrections for the three-body potential were not included.
However previous analyses \cite{sarsa03} have shown 
that they are small for systems with 66 neutrons.

\begin{table}
\caption{\label{tab:bf}
AFDMC energies per particle in MeV for the $AU6'$ and $AU8'$
interactions obtained for a system  of 14 or 66 neutrons in a periodical box
with the guiding functions, $\Psi$, JS and JSB (with $\beta=1$), at $\rho_0$.
Error bars for the last digit are shown in parentheses.
}
\begin{ruledtabular}
\begin{tabular}{ccc}
$\Psi$  & $N=14$      & $N=66$          \\
\hline
JS(AU6')   & $48.27(9)$  & $ 53.11(9)$    \\
JS(AU8')   & $48.4(1) $  & $ 54.4(6) $    \\
JSB(AU8')  & $46.8(1) $  & $ 52.9(2) $
\end{tabular}
\end{ruledtabular}
\end{table}

Fig.~\ref{fig1} shows the AFDMC energies per particle of 14 neutrons
in a periodic simulation cell at $\rho=2\rho_0$ for the $AU6'$ and
$AU8'$ interactions, as a function of the strength parameter $\beta$.
One can see that, for the $AU8'$ interaction the energy minimum is around
$\beta=1$, consistent with the conjecture that $f_b(r)$ moves the nodes
in the optimal way when the spin--orbit potential is included 
in the Hamiltonian. We have obtained a lowering of the energy of
1.6 MeV
with respect to the case of the JS nodal surface, which corresponds to
$\beta=0$ in the figure. If we switch off the spin--orbit potential
by considering the $AU6'$ interaction, the minimum is found at
$\beta=0$, confirming that the spin--backflow nodes are energetically 
advantageous only in the presence of the spin--orbit component of $v_{8'}$. 

AFDMC simulations for the $AU8'$ Hamiltonian with the JSB guiding function
have been also carried out for 66 neutrons. As in the case of the 14
neutron system, we have found a minimum of the energy at $\beta\sim 1$.
The result obtained is compared in Table~\ref{tab:bf}  
with those at $\beta=0$ for both the $AU8'$ and the $AU6'$ interactions.
In spite of sizable
differences between the energies per particle of the 14 neutron
and the 66 neutron systems, the gain in energy $(E(JSB)-E(JS))/N$ 
is roughly independent on the number of particles in the box.
The large differences between
E(14) and E(66) are mainly due to the effect of three--body
interaction. It has been shown \cite{sarsa03} that the finite size effects on 
E(66) are rather small.

The dependence of the JSB energy on the density is reported in
Table~\ref{tab:density} for the 14 neutron system. There is a very
weak dependence of $\Delta E_{LS}$ on the density, in contrast with
VCS results.

In order to make comparison with  recent quantum Monte Carlo calculations 
performed for 14 neutrons interacting via the $v_8'$ 
two--body potential by \textcite{carlson03},
we report in Table~\ref{tab:carlson} the corresponding AFDMC results.
The table displays the energy difference $\Delta E_{LS}$ between the energy
obtained with $v_8'$ and $v_6'$ two--body potential. 

\begin{table}
\caption{\label{tab:density}
AFDMC energies per particle in MeV for the $AU6'$ and $AU8'$
interactions obtained for a system  of 14  neutrons in a periodical box
with the guiding functions, $\Psi$, JS and JSB (with $\beta=1$), 
as a function of the density $\rho$. 
Error bars for the last digit are shown in parentheses.
}
\begin{ruledtabular}
\begin{tabular}{cccc}
$\rho$  & $AU6'$      & $JS$      & $JSB$      \\
\hline
$\rho_0$   & $19.73(5)$  & $19.76(6)$   & $18.76(5)$   \\
$2\rho_0$  & $48.27(9) $ & $48.4(1) $   & $46.8(1)$    
\end{tabular}
\end{ruledtabular}
\end{table}

The calculations of Ref.~\cite{carlson03}, to which the results reported 
in Table~\ref{tab:carlson} refer, have been performed with the $v_{8'}$
potential cut off at the edge of the box. Therefore, the values of
$\Delta E_{LS}$ extracted from there might not be completely comparable
with ours, since ours are already tail corrected and have been 
obtained without introducing any discontinuity in the potential. 
AFDMC seems to agree reasonably well with GFMC in the constrained
path approximation (GFMC--CP). The VCS estimate seems to be too large.

\section{Conclusions}
\label{sec.5}

\begin{table}
\caption{\label{tab:carlson}
Spin--orbit contribution to the energy per particle in MeV of neutron matter
at density $\rho_0$.
The constrained (CP) and unconstrained (UC) GFMC results, as well as the
VCS ones are taken from Ref.~\cite{carlson03}. The AFDMC results
obtained with the JS guiding function are taken from Ref.~\cite{sarsa03}.
Error bars for the last digit are shown in parentheses.
}
\begin{ruledtabular}
\begin{tabular}{cc}
method      &  $\Delta E_{LS}$    \\
\hline
GFMC--CP    &  -1.26(4)  \\
GFMC--UC    &  -2.9(3)    \\
AFDMC--JS   &  -0.14(6)   \\
AFDMC--JSB  &  -1.2(1)    \\
VCS         &  -3.8
\end{tabular}
\end{ruledtabular}
\end{table}

In this paper we have proposed a new kind of space--spin 
orbitals with a spin--backflow form,
which is particularly useful for taking into account
the spin--orbit interaction of nuclear systems in quantum Monte
Carlo simulations. An efficient parameterization of the
spin--backflow is obtained from the spin--orbit correlation of
FHNC/SOC theory. These spin-backflow orbitals can be conveniently used
also in other quantum Monte Carlo calculations, for instance
the Green's Function Monte Carlo simulations of small nucleon systems. 
The nodal surface provided by this new guiding function
is able to decrease the energy about 5 per cent. This amount is not
sufficient to solve the spin--orbit discrepancy between the variational
chain summation and the AFDMC results, however the AFDMC results
are in good agreement
with constrained GFMC simulations.

\begin{acknowledgments}
SAV wants to thanks the kind hospitality of SF and SISSA where most of
this work has been accomplished. 
Portions of this work were supported by the Italian  MIUR-National Research
Project 2001025498.
AS acknowledges the Spanish Ministerio 
de Ciencia y Tecnologia for partial support under contract BMF2002-00200
\end{acknowledgments}

\bibliography{bf}
\bibliographystyle{apsrev}

\end{document}